\documentclass[twoside]{ilcws07}
\usepackage[latin1]{inputenc}
\usepackage[dvips]{graphicx,epsfig,color}
\usepackage{wrapfig,rotating}
\usepackage{amssymb,amsmath,array}

\begin{document}
\title{
%%%%   Paper title goes here  %%%%%%%%%%%%%%
Status reports from the GRACE Group
} %%
%***********************************************************************
% AUTHORS INFORMATION AREA
%***********************************************************************
\author{
Y. Yasui$^1$,
T. Ueda$^2$,
E. de Doncker$^3$,
J. Fujimoto$^2$,\\
N. Hamaguchi$^4$,
T. Ishikawa$^2$,
Y. Shimizu$^5$ 
and 
F. Yuasa$^2$
\\
\vspace{.3cm}\\
% Addresses and institutions (remove "1- " in case of a single institution)
1- Tokyo Management College\\
Ichikawa, Chiba 272-0001, Japan
\vspace{.1cm}\\
2- High Energy Accelerator Research Organization (KEK)\\
1-1 OHO Tsukuba, Ibaraki  305-0801, Japan
\vspace{.1cm}\\
3- Western Michigan University\\ 
Kalamazoo, MI 49008-5371, USA
\vspace{.1cm}\\
4- Hitachi, Ltd., Software Division\\
Totsuka-ku, Yokohama, 244-0801, Japan
\vspace{.1cm}\\
5- The Graduate University for Advanced Studies, Sokendai\\
Shonan Village, Hayama, Kanagawa 240-0193, Japan
}
%%**********************************************************************
% END OF AUTHORS INFORMATION AREA
%***********************************************************************

\maketitle

\begin{abstract}

We discuss a new approach for the numerical 
evaluation of loop integrals. 
The fully numerical 
calculations of an infrared one-loop vertex and a box diagram 
are demonstrated. 
To perform these calculations, we apply an extrapolation method
based on the $\epsilon$-algorithm. 
In our approach, the super high precision control 
in the numerical manipulation 
is essential to handle the infrared singularity. 
We adopt a multi-precision library named {\tt HMLib} 
for the precision control in the calculations. 
\end{abstract}

\section{Introduction}

Measurements of fundamental parameters with high precision will be 
one of the important issues at the ILC experiment. 
We suppose to determine the masses and couplings for 
the standard model (SM) and some of the 
beyond the standard models like the minimal supersymmetric model(MSSM) 
within a few percent precision~\cite{ILCTDR}. 
These precision measurements require the knowledge of 
high precision theoretical predictions, especially
the computation of higher loop corrections. 
Since,the vast number of Feynman diagrams appear in the  
loop calculations, 
performing such computation is absolutely beyond 
the human power  if it should be done by hand.
The procedure of a perturbation 
calculation is well established, thus computers must be able 
to take the place of human hand. 
In this purpose several groups have developed computer programs which 
generate Feynman diagrams and calculate cross sections automatically, 
like {\tt GRACE}~\cite{grace},
{\tt FeynArts-FormCalc}~\cite{THahns} and {\tt CompHEP}~\cite{comps}.

The {\tt GRACE} system, developed by Minamitateya group, 
is one of the systems to calculate Feynman amplitudes 
including loop diagrams.
Our final goal of the {\tt GRACE} system is to construct 
the fully automatic computation system of multi-loop integrals.
In this stage, it is successfully working  
for one-loop calculations in both the SM~\cite{grc}
and the MSSM~\cite{gsusy}. 
However, the fully automatic way to compute multi-loop integrals  
is not established, thus the system is still semi-automatic. 
Especially, since infrared singularities require the 
analytic manipulations,  
it makes difficult to handle the singularity in the automatic way. 
To avoid such a difficulty, we really need to treat loop integrals 
in a fully numerical way.

In this paper, we discuss the fully numerical computation 
of one-loop integrals with existing infrared singularities.
First, we demonstrate the calculations of an one-loop vertex 
and a box diagram in QED case. 
We put a fictitious photon mass as a regulator of the infrared 
singularities. 
For the calculations, we adopt the very brute force way 
as the numerical extrapolation method. In our approach, 
the $\epsilon$-algorithm is efficient. 
We also introduce a precision control technique to handle 
the numerical calculation with extremely small photon masses.

In QCD, one can not avoid to use the dimensional regularization
to treat the infrared singularity. The sector decomposition
technique \cite{sd} is useful to pick up the singularities but in general,
the numerical integration of each term is not trivial because
the denominators of the integrands can be zero in the integral
region. Therefore, the extrapolation method should be effective
in this case, too.

The layout of this paper is as follows. 
In section 2, we show the basic idea and formula of the loop
integrals.
Numerical results of the infrared one-loop vertex 
and box diagram are shown in section 3. 
The case for the dimensional regularization is discussed in section 4. 
In section 5 we will summarize this paper.

%---------------------------------------------------------
%\begin{wraptable}{l}{0.6\columnwidth}
\begin{table}[b]
\centerline{
\begin{tabular}{|l|r|r|} \hline
$\lambda$ [GeV] & Average lost-bits & Maximum lost-bits\\ \hline 
$10^{-20}$ & 88 & 92\\ \hline
$10^{-21}$ & 98 & 102\\ \hline
$10^{-22}$ & 108 & 112\\ \hline
\end{tabular}}
\caption{Lost-bits information from {\tt HMLib} with 
the quadruple precision arithmetic. $\lambda$ is a fictitious photon mass.}
\label{tab:hmlib}
%\end{wraptable}
\end{table}
%---------------------------------------------------------

\section{Basic idea}

The basic idea of our numerical method is the combination of an efficient
multi-dimensional integration routine and an extrapolation method.
Here, we consider a scalar one-loop n-point integral given by
\begin{equation}
{I}(\epsilon) 
= \int \frac{d^4 l}{(2 \pi)^4 i} \frac{1}{(l^2 - m_1^2 + i
   \epsilon) ((l + p_1)^2 - m_2^2 + i \epsilon)
   \cdots ((l + \sum_{j =
   1}^{n - 1} p_j)^2 - m_n^2 + i  \epsilon)},
\end{equation}
%-------------------------------------------------------
\begin{wrapfigure}{r}{0.3\columnwidth}
%\begin{figure}
\centerline{
\includegraphics[width=0.25\columnwidth,]{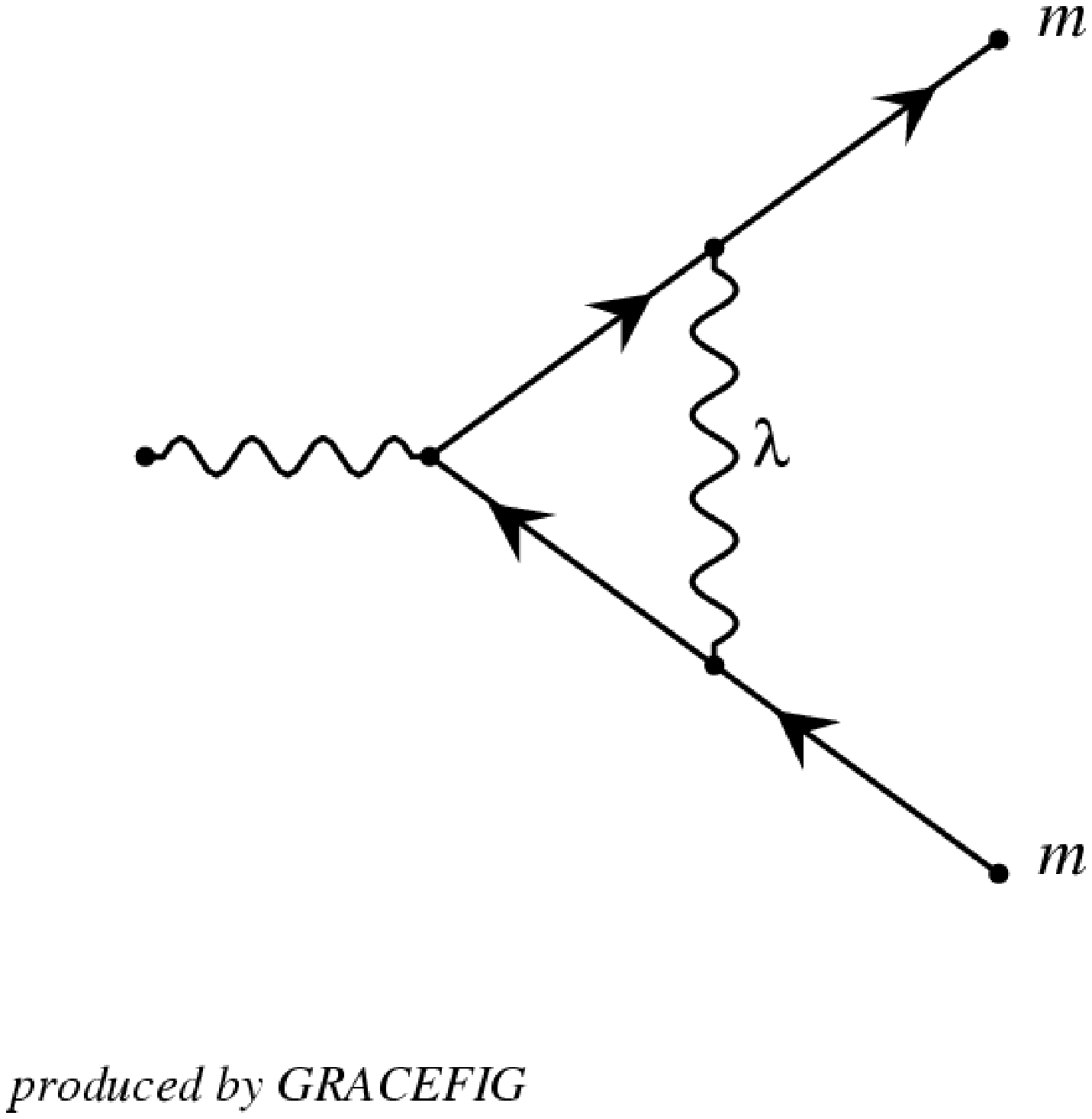}
}
\caption{One-loop vertex diagram}
\label{Fig:vertex}
\end{wrapfigure}
%-------------------------------------------------------
where $l$ is the loop momentum, $p_j$ is the momentum 
of the $j$th external particle and $m_j$ is the mass carried by 
the $j$th internal line. Putting $\epsilon > 0$ in the 
denominator of Feynman integrals to prevent the integral 
from diverging, we can get the numerical results of ${I}(\epsilon)$ 
for a given $\epsilon$. 
Calculating a sequence of ${I}(\epsilon_l)$ 
with various value of $\epsilon_l=\epsilon_0\times (const.)^{-l}$ 
$(l=0,1,2,\cdots)$ and extrapolating into the limit of 
$\epsilon\rightarrow 0$, 
we can get the final result of the integration
which appears in a physical amplitude.

In this approach, we use {\tt DQAGE} routine in the 
multi-dimensional integration.
It is included in the package {\tt QUADPACK} \cite{QUADPACK} and it is
a globally adaptive integration routine.
We also apply the $\epsilon$-algorithm introduced by
P.Wynn\cite{wynn}.
The algorithm accelerates the conversion of the sequences.
Our approach is very efficient for massive one-loop and two-loop 
integrals including non-scalar type integrals~\cite{dq1, dq2, dq3,
dq4}.

In the above massive cases, 
the results are consistent with analytic ones very well, even in the 
double precision arithmetic or quadruple precision arithmetic. 
On the other hand, for massless cases, 
we need much more careful treatments to handle the
infrared singularities. For the infrared divergent diagram,  
it becomes harder to get a result with an enough accuracy
even in the quadruple precision arithmetic.
Thus, the precision control of the numerical integration 
becomes another important elements in our approach~\cite{acat07}.

%-------------------------------------------------------
\begin{wrapfigure}{r}{0.29\columnwidth}
\centerline{
\includegraphics[width=0.27\columnwidth,
height=4cm]{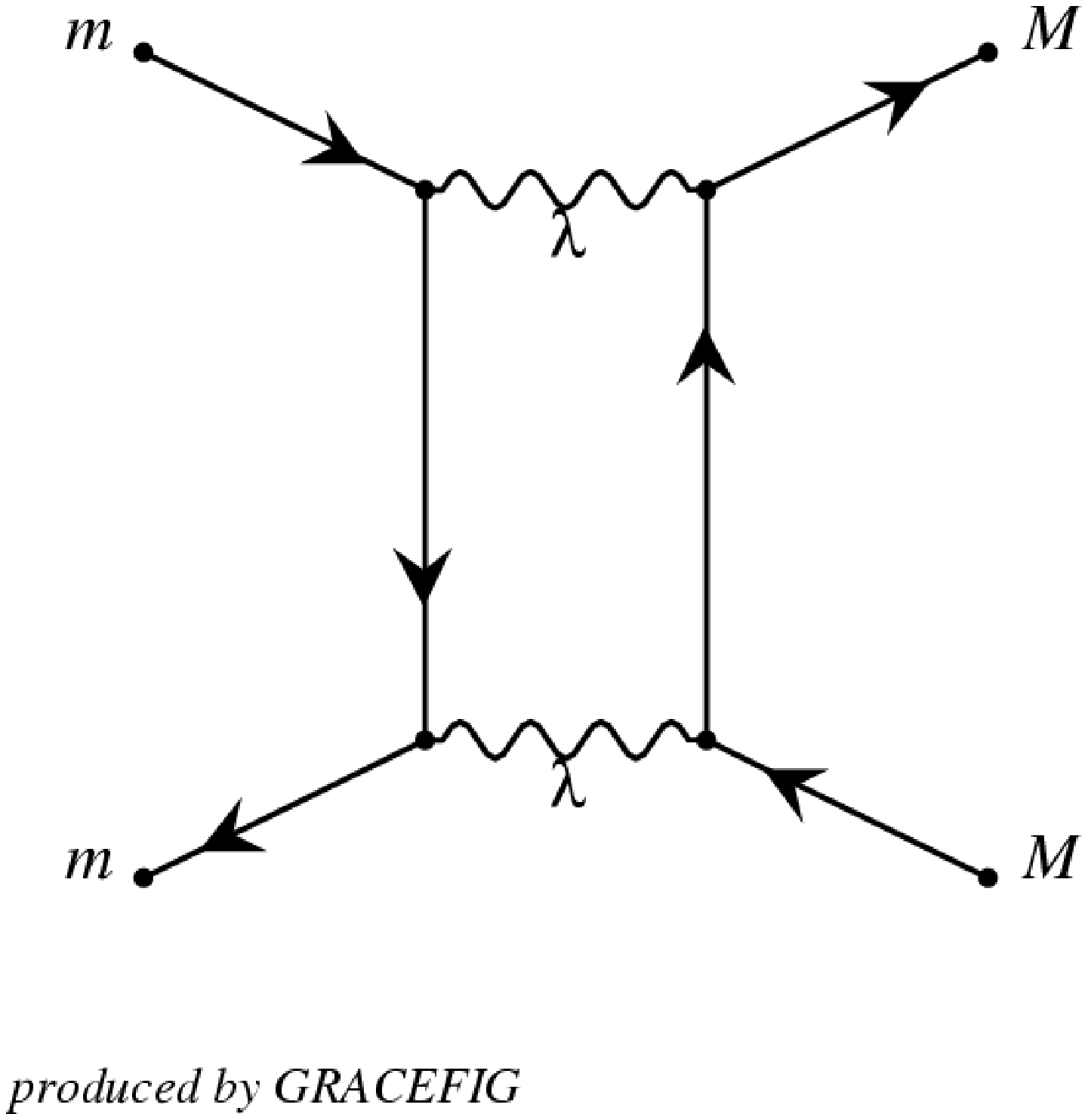}
}
\caption{One-loop box diagram in QED}
\label{Fig:box}
%\end{figure}
\end{wrapfigure}
%------------------------------------------------------- 
For the precision control, we use 
{\tt HMlib}\cite{hmlib} as the multi-precision library. 
The advantage of this library is that it gives an information of 
the lost-bits during the calculation, 
thus we can guarantee the precision of the results. 
In Table~\ref{tab:hmlib},  we show an example of the lost-bits
information supplied by {\tt HMLib}.
Here, we demonstrated the calculation of one-loop vertex diagram
(Fig.\ref{Fig:vertex}).  
The diagram includes the infrared singularity due to the photon 
exchange. We put a fictitious photon mass $\lambda$ 
as a regulator of the infrared singularity.
In {\it P}-precision presentation implemented in
{\tt HMlib}, a sign bit is 1 bit and an exponent bit is 15 bits and 
a mantissa is ($32 \times P - 16$) bits with based on IEEE754 FP. 
When $P = 4$ (quadruple precision), the mantissa becomes 112 bits. 
In the Table when $\lambda$ is $10^{-23}$ GeV, the maximum number 
of lost-bits is the same as the number of bits of the mantissa.
This means quadruple-precision is not enough and 
we have to perform the calculation within octuple or higher precision.

\section{Numerical results}

For the one-loop vertex diagram shown in 
Fig.~\ref{Fig:vertex}, the loop integral we consider in this paper is
\begin{equation}
I = \int_{0}^{1} dx \int_{0}^{1-x} dy
{\frac{1}{D}},
\label{eqn:I-vertex}
\end{equation}
where
\begin{equation}
D = -xys + {(x + y)} ^2m^2 + (1 - x - y) \lambda^2.
\label{eqn:D-vertex}
\end{equation}
Here $s$ denotes squared central mass energy and $m$ and $\lambda$
are a mass of external particles and a fictitious photon mass respectively.
Replacing $s$ by $s +i \epsilon$ in (\ref{eqn:D-vertex}) 
and applying extrapolation method, we obtain the 
numerical results which are shown in Table~\ref{tab:E-PT}~\cite{acat07}. 
Results are compared to analytic results evaluated by the formula in
\cite{1loop3}.
The combination of the extrapolation method and multi precision 
control with {\tt HMLib} works very well and we get stable results 
even though the photon mass becomes much smaller as $10^{-160}$ GeV. 

%-----------------------------------------------------------------------
\begin{table}
%\scriptsize{
\centerline{
\begin{tabular}{|l|r|r|r|} \hline
$n$   & Numerical Results & P & Analytical Results [P=4] \\ \hline
{-30} & -0.150899286980769753D-01 $\pm$ 0.771D-26&8&  -0.150899286980482291E-01 \\
      &  0.189229839615898822D-02 $\pm$ 0.124D-25&8&   0.189229839615525389E-02 \\ \hline
{-80} & -0.405390396284235075D-01 $\pm$ 0.580D-15&16& -0.405390396283445396E-01 \\
      &  0.478581216125478532D-02 $\pm$ 0.401D-12&16&  0.478581216112143981E-02 \\ \hline
{-120}& -0.608983283726997427D-01 $\pm$ 0.556D-15&32& -0.608983283725815879E-01 \\
      &  0.710062317325786663D-02 $\pm$ 0.415D-12&32&  0.710062317309438855E-02 \\ \hline
{-160}& -0.812576170752810666D-01 $\pm$ 0.549D-10&32& -0.812576183472699269E-01 \\
      &  0.941543418501223556D-02 $\pm$ 0.109D-11&32&  0.941543432496722442E-02 \\ \hline
\end{tabular}}
%}
\caption{Numerical results of the one-loop vertex diagram 
with $\sqrt{s} =
500$ GeV, 
$t = -150^{2}$ $GeV^{2}$, $m = m_{e} = 0.5 \times 10^{-3}$ GeV.
Photon mas $\lambda = 10^n$[GeV] and P denotes P-precision. 
The upper is the result of Real part and the lower is one of Imaginary part.}
\label{tab:E-PT}
\end{table}

We also demonstrate the calculation of the one-loop box diagram shown in 
Fig.\ref{Fig:box}.
Here, we consider the following integral, 
\begin{equation}
I = \int_{0}^{1} dx \int_{0}^{1-x} dy \int_{0}^{1-x-y} dz
{\frac{1}{D^2}},
\label{eqn:I-box}
\end{equation}
where
\begin{eqnarray}
D &=& -xys - tz(1-x-y-z)+ (x+y)\lambda^2 \\
&+& (1-x-y-z)(1-x-y)m^2 +z(1-x-y)M^2.
\label{eqn:D-box}
\end{eqnarray}
$s$ denotes squared central mass energy and 
$m$ and $M$ are external masses of particles. 
Again, replacing $s$ by $s +i \epsilon$ in 
eq.(\ref{eqn:D-box}) 
and applying extrapolation method, we obtain the 
numerical results.
We show the real part of the numerical results  
with the analytic ones 
in Table~\ref{tab:gg-PT}
as an example~\cite{acat07}. 
The results are almost consistent each other.
However, 
since high precision computation costs huge CPU time, 
we only have done the quadruple precision arithmetic.
Thus the results are not reliable when the
photon mass is smaller as $10^{-30}$ GeV.

%--------------------------------------------------------
\section{Dimensional Regularization}
%-------------------------------------------------------
\begin{wrapfigure}{r}{0.22\columnwidth}
\centerline{\includegraphics[width=0.15\columnwidth]{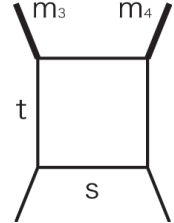}}
\caption{One-loop box diagram with two massive external legs, 
so called "hard" case.}
\label{Fig:SD-box}
\end{wrapfigure}
%-------------------------------------------------------
We next consider the diagram contains two massive and two massless 
external legs, and internal lines are all massless with the dimensional
regularization, where $D=\varepsilon+2$ (Fig. \ref{Fig:SD-box}). 
In order to pick up the singularities, 
we apply the sector decomposition technique.
The loop integral is expressed as follows:
\begin{eqnarray}
I_4 &=& \int \prod_{i=1}^{4} dx_i {\frac{\delta(1-\sum x_i)}
{(-sx_1x_3-tx_2x_4-m_3^2x_3x_4-m_4^2x_1x_4)^{2-\varepsilon}}}, \\
&=& \sum_{n=-2,-1,0,\cdots}C_n \times \varepsilon^n.
\label{eqn:SD-box}
\end{eqnarray}
In the case of $s=123$ GeV$^2$,$t=-200$ GeV$^2$, 
$m_3^2=50$ GeV$^2$ and $m_4^2=60$ GeV$^2$, 
the denominator of each integrand 
corresponding to $1/\varepsilon$ and the finite term can be zero. 
Again, $+i \epsilon$ is introduced in the denominator of
eq.(\ref{eqn:SD-box}) 
to specify the analyticity of it
and the extrapolation method does work to perform the numerical integration. 
Table \ref{tab:SD-box} shows the complete numerical integration 
can reconstruct the results with
the analytical expressions~\cite{analytic}.  

%----------------------------------------------------
\begin{table}
%\scriptsize{
\centerline{
\begin{tabular}{|l|r|r|r|} \hline 
$n$& Numerical Results & P & Numerical Results [P=4] \\ \hline
{-15}& -0.1927861102670278D-06 $\pm$ 0.314D-14 &4,4,2& -0.1927861122439641E-06 \\ \hline
{-20}& -0.2472486348234972D-06 $\pm$ 0.586D-15 &4,4,2& -0.2472486352599175E-06 \\ \hline
{-25}& -0.3017111253761463D-06 $\pm$ 0.111D-13 &4,4,2& -0.3017111582758710E-06 \\ \hline
{-30}& -0.3562028882831722D-06 $\pm$ 0.867D-10 &4,4,2& -0.3561736812918245E-06 \\ \hline
\end{tabular}}
%}
\caption{Numerical results of the loop integral of one-loop box
diagram with 
$\sqrt{s} = 500$ GeV, $t$ = -$150^2 GeV^2$, $m$ = 0.5 $\times 10^{-5}$  
GeV, $M$ = 150 GeV. 
Photon mas $\lambda = 10^n$[GeV] and P denotes P-precision. 
This is results of Real part.}
\label{tab:gg-PT}
\end{table}

%----------------------------------------------------
\begin{table}
%\scriptsize{
\centerline{
\begin{tabular}{|l|r|r|r|} \hline 
$n$& Numerical Results& Analytic Results       \\ \hline
{-2}& -0.40650406505E-04 &-0.40650406504E-04   \\ \hline
{-1}& -0.34156307031E-03 &-0.34156306995E-03   \\ \hline
{0} & -0.14929502492E-02 &-0.14929502456E-02   \\ \hline
\end{tabular}}
%}
\caption{Numerical v.s. Analytical results in the double 
precision arithmetic. 
Real part of $C_n$ of Fig. \ref{Fig:SD-box} with 
$s$ = 123 GeV$^2$, $t$ = -200 GeV$^2$, 
$m_3^2$ = 50 GeV$^2$ and $m_4^2$=60 GeV$^2$. }
\label{tab:SD-box}
\end{table}

\section{Summary}
We discussed a new approach of the numerical method for 
multi-loop integrals. 
The combination of an efficient
multi-dimensional integration routine and an extrapolation method 
is applicable to the one-loop integral even though it includes 
infrared singularities both with the introduction of the 
fictitious mass and the dimensional
regularization.
We demonstrated the fully numerical calculations for 
one-loop vertex and box diagram with infrared singularities.
To handle the infrared singularities, 
high precision control is essential and 
in some case quadruple precision is not enough to 
obtain the stable results. 
In our calculations, we applied the multi-precision library 
{\tt HMLib} which has the high performance of the precision 
control .
Our method is efficient and we obtain reliable results
for one-loop vertex and box diagrams with infrared singularities.

\section*{Acknowledgments}
We wish to thank Prof. Kaneko and Dr. Kurihara 
for their valuable suggestions. 
We also wish to thank Prof. Kawabata for his encouragement and support.
This work was supported in part by the Grants-in-Aid 
(No.17340085) of JSPS.
This work was also supported in part by the ``International
Research Group'' on ``Automatic Computational Particle Physics''
(IRG ACPP) funded by CERN/Universities in France,
KEK/MEXT in Japan and MSU/RAS/MFBR in Russia.

%\section*{Appendix A}

% ****************************************************************************
% BIBLIOGRAPHY AREA
% ****************************************************************************

%\begin{footnotesize}
% IF YOU DO NOT USE BIBTEX, USE THE FOLLOWING SAMPLE SCHEME FOR THE REFERENCES
% ----------------------------------------------------------------------------

% ----------------------------------------------------------------------------

\end{document}